\documentclass[article]{jss}

\usepackage{thumbpdf,lmodern}

\usepackage{framed}

\usepackage{subfig}

\usepackage{amsmath}

\author{Alex Fisch\\Lancaster University
  \And Daniel Grose\\Lancaster University
  \And Idris A. Eckley\\Lancaster University
  \AND Paul Fearnhead\\Lancaster University 
   \And Lawrence Bardwell\\Lancaster University}
\Plainauthor{Alex Fisch, Daniel Grose, Idris A. Eckley, Paul Fearnhead, Lawrence Bardwell}

\title{\pkg{anomaly} : Detection of Anomalous Structure in Time Series Data}
\Plaintitle{anomaly : Detection of Anomalous Structure in Time Series Data}
\Shorttitle{Detection of Anomalous Structure in Time Series Data}

\Abstract{
One of the contemporary challenges in anomaly detection is the ability to detect, and differentiate between, both point and collective anomalies within a data sequence or time series. The \pkg{anomaly} package has been developed to provide users with a choice of anomaly detection methods and, in particular, provides an implementation of the recently proposed Collective And Point Anomaly family of anomaly detection algorithms. This article describes the methods implemented whilst also highlighting their application to simulated data as well as real data examples contained in the package. 
}


\Address{
  Daniel Grose\\
  Department of Mathematics and Statistics \\
  Fylde College \\
  Lancaster University\\
  LA1 4YF \\
  United Kingdom \\
  E-mail: \email{dan.grose@lancaster.ac.uk}\\
  URL: \url{https://www.lancaster.ac.uk/sci-tech/about-us/people/daniel-grose}
}

\begin{document}



\section[Introduction]{Introduction} \label{section:introduction}

Within this article, we focus on the challenge of detecting anomalies within data sequences. Anomaly detection has become an increasingly important area of research activity due to its wide ranging application: from fault detection \citep{theissler2017detecting,zhao2018anomaly}, to fraud
prevention \citep{ahmed2016survey}, and system monitoring \citep{goh2017anomaly}. In broad terms, anomalies are observations that do not conform with the general or local pattern of the data and are commonly considered to fall into one of three categories: global anomalies, contextual anomalies, or collective anomalies \citep{nchandolaanomaly-detection-overview}. 
Global anomalies and contextual anomalies are defined as single observations that are outliers with regards to the
complete dataset and their local context respectively. Conversely, collective anomalies are defined as sequences
of observations that are not anomalous when considered individually, but together form an anomalous pattern \citep{2018arXiv180601947F}.
 
In parallel with the methodological development of statistical anomaly detection for data sequences, a number of software implementations have been developed. For example, within \proglang{R}, the \pkg{anomalize} package \citep{anomalize-package} provides an implementation of two point anomaly approaches, based on the interquartile range and generalized extreme studentized deviate test respectively, following the removal of any seasonal and trend components. Similarly \pkg{otsad} \citep{otsad-package} also provides a suite of approaches for the detection of point anomalies, whilst \pkg{cbar} \citep{cbar-package} seeks to identify contextual anomalies using a Bayesian framework. Conversely, \pkg{tsoutliers} \citep{tsoutliers-package} seeks to detect innovative and additive outliers together within time series, whilst \pkg{oddstream} \citep{oddstream-package} implements an algorithm for the detection of anomalous series within newly arrived collections of series and \pkg{stray} \citep{stray-package} implements the HDoutliers algorithm for various settings including the detection of anomalies in high-dimensional data. Whilst the aforementioned packages arguably represent the current state of the statistical art at the time of writing, a number of other contributions have been made by researchers in other disciplines: see for example Python packages including \pkg{anomatools} \citep{anomatools-package}, \pkg{adtk} \citep{adtk-package} and \pkg{PySAD} \cite{pysad-package}, and Julia contributions including \pkg{MultivariateAnomalies} \citep{esd-8-677-2017} and \pkg{AnomalyDetection} \citep{anomalydetection-package-julia}.



This paper describes the \pkg{anomaly} package \citep{anomaly-package} that implements a number of recently proposed methods for anomaly detection. For univariate data there is the Collective And Point Anomaly detection (CAPA) method of \cite{2018arXiv180601947F}, that can detect both collective and point anomalies. For multivariate data there are three methods, a multivariate extension of CAPA \citep{2019arXiv190901691F}, the Proportion Adaptive Segment Selection (PASS) method of \cite{10.1093-biomet-ass059}, and a Bayesian approach, Bayesian Abnormal Region Detector \citep{bardwell2017}.

The multivariate CAPA method and PASS are similar in that, for a given segment they use a likelihood-based approach to measure the evidence that it is anomalous for each component of the multivariate data stream, and then merge this evidence across components. They differ in how they merge this evidence, with PASS using higher criticism \citep{donoho2004} and CAPA using a penalised likelihood approach. One disadvantage of the higher criticism approach for merging evidence is that it can lose power when only one or a very small number of components are anomalous. Furthermore, CAPA also allows for point anomalies in otherwise normal segments of data, and can be more robust to detecting collective anomalies when there are point anomalies in the data. CAPA can also allow for the anomalies segments to be slightly mis-aligned across different components. 

The BARD method considers a similar model to that of CAPA or PASS, but is Bayesian and so its basic output are samples from the posterior distribution for where the collective anomalies are, and which components are anomalous. It does not allow for point anomalies. As with any Bayesian method, it requires the user to specify suitable priors, but the output is more flexible, and can more directly allow for quantifying uncertainty about the anomalies.



The article begins by providing a brief introduction to anomaly detection before proceeding to give a detailed treatment of each approach. In each case, the relevant methodology is introduced, describing
the associated package functionality where appropriate. The methods are applied to a number of test datasets that are available with the package. These data sets comprise the machine temperature data introduced by \cite{DBLP:journals/corr/LavinA15}, and a microarray genomics dataset. The examples also include details of how the effects of autocorrelation can be accounted for through the adjustment of the method parameters or by applying transforms to preprocess the data prior to analysis.  

\section[Background]{Background} \label{section:Background}

The suite of methods described in this article focuses on collective anomalies. 
Informally, collective anomalies are segments of data which are anomalous when compared against the general structure of the full data. The modelling paradigm is to assume that there is a common model for data outside the anomalous regions, for example that it is independent normally distributed with a fixed mean and variance, and that collective anomalies correspond to segments of the data that are inconsistent with this, for example due to having a different mean or variance. 
One approach to modelling this type of anomaly is via epidemic changepoints -- a particular form of changepoints admitting one change away from the typical distribution of the data and one back to it at a later time \citep{2018arXiv180601947F}. Formally, in the univariate setting, data, $\{ {x}_{t}\}$, are said to follow a parametric epidemic
changepoint model if $\{{ x}_{t}\}$ obey the parametric model  $f(x_t,{\boldsymbol \theta}(t))$ at all times and the parameter ${\boldsymbol \theta}(t)$ satisfies
\begin{align}\label{eq:univariate}
\theta(t) = \begin{cases}
\theta_1  & s_1 < t \leq e_1, \\
&\vdots \\
\theta_K  & s_K < t \leq e_K, \\
\theta_0  & \text{otherwise}.
\end{cases}
\end{align}
Here $(s_1,e_1),....,(s_K,e_K)$ denote the start and end points of $K$ collective anomalies. The typical (baseline) behaviour of the data sequence is defined by the parameter $\theta_0$. Conditionally on the parameter $\theta(t)$, all observations ${x}_{t}$ are assumed to be independent, with relaxations of this assumption being discussed in the following sections.

When extending to the multivariate setting, i.e., a $p$-dimensional multivariate time series, it is common to assume that the series are independent, but that their periods of anomalous behaviour align. The copy number variations data set \citep{2011arXiv1106.4199B} provides a good example of such behaviour. In the absence of a copy number variation, data from different individuals can be assumed to be independent. However, when collective anomalies under the form of copy number variations occur, they typically affect a subset of the test subjects. Under such a model, it is well known that joint analysis can lead to significant improvements in detection power over analysing each component individually \citep{donoho2004}. The subset multivariate epidemic changepoint model provides a natural model for this type of behaviour. It assumes that 
\begin{equation}\label{eq:multivariate}
\boldsymbol{\theta}^{(i)}(t) = \begin{cases}
\boldsymbol{\theta}^{(i)}_1  & \text{if} \; s_1  < t \leq e_1   \; \text{and} \; i \in \textbf{J}_1, \\
&\vdots \\
\boldsymbol{\theta}^{(i)}_K  & \text{if} \; s_K  < t \leq e_K \; \text{and} \; i \in \textbf{J}_K,\\
\boldsymbol{\theta}^{(i)}_0  & \text{otherwise},
\end{cases} \;\;\;\;\;\;\;\;\ 1 \leq i \leq p.
\end{equation}
where, again, $K$ is the number of collective anomalies with $(s_k$,$e_k)$ denoting the start and end of the $k$th collective anomaly. The $k$th collective anomaly only affects 
the subset $J_k$ of time-series. If the $i$th time-series is affected by the $k$th collective anomaly, i.e., $i \in J_k$ then $\theta_k^{(i)}$ denotes its parameter 
value; with $\theta_0^{(i)}$ denoting the parameter governing the typical behaviour of the $i$th time-series.

\section[CAPA-family]{The Collective And Point Anomaly Family} \label{section:capa}

The Collective And Point Anomaly (CAPA) family of algorithms \citep{2018arXiv180601947F,2019arXiv190901691F} differ from many other anomaly detection methods in that they seek to simultaneously detect and
distinguish between both collective and point anomalies. 
CAPA assumes that the data $\{{\bf x}_{t}\}$ follow the model detailed in \eqref{eq:univariate}, when 
univariate or \eqref{eq:multivariate} when 
multivariate. Point anomalies are incorporated within the model as epidemic changes of length one. When analysing multivariate data, CAPA assumes that the collective anomalies don't overlap, i.e., that $e_1 \leq s_2, ... , e_{K-1} \leq s_K$, whilst allowing for the alignment of collective anomalies to be imperfect, i.e., allowing the components to leave their typical state and return to it at slightly different times.

Whilst the CAPA procedure can allow for many different models for the data, the current implementation assumes that the data is independent and normally distributed, and that the data has been normalised so that the mean is 0 and variance is 1. Non-anomalous data points are drawn from a normal distribution with a specific mean and variance (that the CAPA algorithm will estimate). Collective anomalies correspond to regions where the mean or mean and variance of the data are different. 

CAPA infers the number, $K$, and locations $(s_1,e_1),...,(s_K,e_K)$ of collective anomalies as well as the set $O$ of point anomalies by maximising the penalised saving function
\begin{equation}\label{eq:Saving}
\sum_{i=1}^K \left( S(s_i,e_i) - \beta(e_i-s_i) \right) + \sum_{t \in O} \left( S'(x_t) - \beta' \right),
\end{equation}
with respect to $(s_1,e_1),...,(s_K,e_K)$ and $K$, subject to constraints on the maximum and minimum lengths of anomalies (see \cite{2018arXiv180601947F} for details). Here the saving statistic, $S(s,e)$, of a putative anomaly with start point, $s$, and end point, $e$, corresponds to the improvement in model fit obtained 
by modelling the data in segment $(s,e)$ as a collective anomaly. Given this improvement will always be non-negative a penalty, $\beta(e_i-s_i+1)$, potentially depending on the length of the putative anomaly is used to prevent false positives being flagged. The choice of the penalty is model dependent, and discussed in the following sections. Similarly, $S'(x_t)$ and $\beta'$ denote the improvement in model 
fit by assuming observation, $x_{t}$, is a point anomaly. 

CAPA makes some important independence assumptions, and also assumes that the mean and variance of the non-anomalous data is constant. As we see below, it can successfully be applied to situations where these assumptions do not hold. There are two approaches to do so. First we can transform the data so that the assumptions are more reasonable -- this could be to remove the effect of common factors that induce dependence across components or applying a filter to remove auto-correlation from the noise. Alternatively we can inflate the default penalties so that we still have good properties if there are no collective anomalies.
We give an example of this latter approach for the machine temperature data set below.

CAPA maximises the penalised saving in (\ref{eq:Saving}) using an optimal partitioning algorithm
\citep{2005ISPL-short}. By default, the runtime of CAPA family algorithms scales quadratically in the number of observations. In practice, the computational complexity
can be reduced by applying a pruning technique developed by \citet{killick-jasa} that is used in the \pkg{changepoint} package \citep{killick-jss}. It is particularly effective when a large number of anomalies is present -- leading to a linear relationship between runtime and data size when the number of anomalies is proportional to the size of the data. Another way to reduce the runtime is to impose a maximum length, $m$, for anomalies, the runtime then scaling linearly in both the number of observations and $m$.

The \pkg{anomaly} package contain a single function, \code{capa}, for accessing both the univariate and multivariate methods.  It has the following arguments.
\begin{itemize}
\item \code{x} A numeric matrix with n rows and p columns containing the data which is to be inspected. The time series data classes ts, xts, and zoo are also supported.
\item \code{beta} A numeric vector of length p, giving the marginal penalties.  If beta is missing and p = 1 then beta = 3log(n) when the type is "mean" or "robustmean" and beta = 4log(n) otherwise. If p > 1, type ="meanvar" or type = "mean" and max\_lag > 0 it defaults to the penalty regime 2' described in \cite{2018arXiv180601947F}. If p > 1, type = "mean"/"meanvar" and max\_lag = 0 it defaults to the pointwise minimum of the penalty regimes 1, 2, and 3 in \cite{2018arXiv180601947F}.
\item \code{beta\_tilde} A numeric constant indicating the penalty for adding an additional point anomaly. It defaults to 3log(np), where n and p are the data dimensions.
\item \code{type} A string indicating which type of deviations from the baseline are considered. Can be "meanvar" for collective anomalies characterised by joint changes in mean and
variance (the default), "mean" for collective anomalies characterised by changes in mean only, or "robustmean" (only allowed when p = 1) for collective anomalies characterised by changes in mean only which can be polluted by outliers.
\item \code{min\_seg\_len} An integer indicating the minimum length of epidemic changes. It must be at least 2 and defaults to 10.
\item \code{max\_seg\_len} An integer indicating the maximum length of epidemic changes. It must be at least \code{min\_seg\_len} and defaults to Inf. The computational cost of the CAPA algortihm can be reduced by decreasing the value of \code{max\_seg\_len}.  
\item \code{max\_lag} A non-negative integer indicating the maximum start or end lag. Only useful for multivariate data. Default value is 0.
\end{itemize}
When the \code{x} argument to \code{capa} is one dimensional (i.e. a vector or $n \times 1$ array or matrix) the univariate method is used and the \code{max\_lag} argument is ignored, otherwise, the multivariate method is employed. The \code{capa} function returns an S4 object of type \code{capa.class} for which the generic methods
\code{plot} and \code{summary} have been provided.
\subsection[Univariate CAPA]{Univariate CAPA} \label{subsection:uvcapa}
The \pkg{anomaly} package supports univariate CAPA via the \code{capa} function for detecting segments characterised by an anomalous mean or anomalous mean and variance. When investigating segments for an anomalous mean against a typical Gaussian background of mean 0 and variance 1. If we let $f(x;\mu,\sigma)$ denote the density function for a normal random variable with mean $\mu$ and variance $\sigma^2$ evaluated at $x$, then the savings for a collective anomaly are equal to the improvement in log-likelihood by fitting a segment as anomalous. When only the mean of an anomalous segment changes, 
\[
\max_{\mu} \sum_{t=s}^e \left\{ \log f(x_t;\mu,1)-\log f(x_t;0,1) \right\}. 
\]
While the saving for a point anomaly is set to be the saving for a collective anomaly of length 1. This gives
\begin{equation*}
S(s,e) = (e-s+1) \left( \bar{x}_{s:e} \right)^2   \;\;\;\;\;\;  \text{and} \;\;\;\;\;\; S'(x) = x^2
\end{equation*}
where $\bar{x}_{s:e}$ denotes the mean of observations $x_s,...,x_e$. 
Conversely, when investigating segments for an anomalous mean and/or variance, savings is
\[
\max_{\mu,\sigma} \sum_{t=s}^e \left\{ \log f(x_t;\mu,\sigma)-\log f(x_t;0,1) \right\},
\]
with the saving for a point anomaly being that for a change in variance only in a segment of size 1. This gives
\begin{equation*}
S(s,e) = \sum_{t=s}^e x_t^2 - (e-s+1) \left(1 +  \log \left( \frac{ \sum_{t=s}^e \left( x_t - \bar{x}_{s:e} \right)^2}{e-s+1}\right)   \right) 
\;\; , \;\;
S'(x) = x^2 - 1 - \log\left(e^{-\beta'} + x^2\right)
\end{equation*}
Note that the data, \code{x}, requires standardisation using robust estimates for the typical mean (the median) and the typical variance (the median absolute deviation) obtained on the complete data series so that the above cost functions can be used. See \citep{2019arXiv190901691F} for further details.

%

The argument \code{max_seg_len} sets the maximum length of a collective anomaly. It can be used to prevent the detection of weak but long anomalies which typically arise as a result of model misspecification and also to reduce the run time of the CAPA algorithm. It defaults to a value equal to the length of the data series. Care is needed, as if a value is set that is smaller than the size of the actual anomalous regions, then CAPA is likely to fit multiple collective anomalies to such a region.

By default, $\beta = 3\log(n)$ and $\beta = 4\log(n)$ are used for changes in mean and changes in mean and variance respectively, and $\beta'=3\log n$ for all models, as they have been shown to control the number of false positives when all observations are independent and identically distributed (i.i.d.)\ Gaussian \citep{2018arXiv180601947F,2019arXiv190901691F}.  These default parameters have a tendency to return many false positives on structured, i.e., non independent, data. In this case, \code{beta} and \code{beta_tilde} should be inflated whilst keeping their ratio constant. When looking for changes in mean, using 
\begin{equation}
\beta = \tilde{\beta} = 3 \frac{1+\hat{\rho}}{1-\hat{\rho}}\log(n),
\end{equation}
where $\hat{\rho}$ is a robust estimate for the $AR(1)$-autocorrelation often yields good false positive control. For changes in mean and variance,
\[
\beta = 4 \frac{1+\hat{\rho}}{1-\hat{\rho}}\log(n), ~~ \tilde{\beta} = 3 \frac{1+\hat{\rho}}{1-\hat{\rho}}\log(n).
\]
The specific factor is justified theoretically in \citet{LavielleMarc2000LEoa}. 

Alternatively, the data $x_{t}$ can be directly transformed using
\begin{equation}
x_t' =  \sqrt{\frac{1-\hat{\rho}}{1+\hat{\rho}}} \left(\frac{x_t - \hat{\mu}}{\hat{\sigma}}\right),
\label{eqn:ac_corrected_transform}
\end{equation}
where $\hat{\mu}$ is the median, and $\hat{\sigma}$ is a robust estimator of the standard deviation of the data \code{x}, such as based on the inter-quartile range, or the median absolute deviation from the median. This transform should only be used when looking for mean anomalies.

\subsubsection{Simulated data}  
\label{subsubsection:example-1-uvcapa-simulated-data}
\begin{figure}
\begin{minipage}{.5\linewidth}
\centering
\subfloat[]{\label{figure:uvcapa-simulated-data-1}\includegraphics[scale=.5]{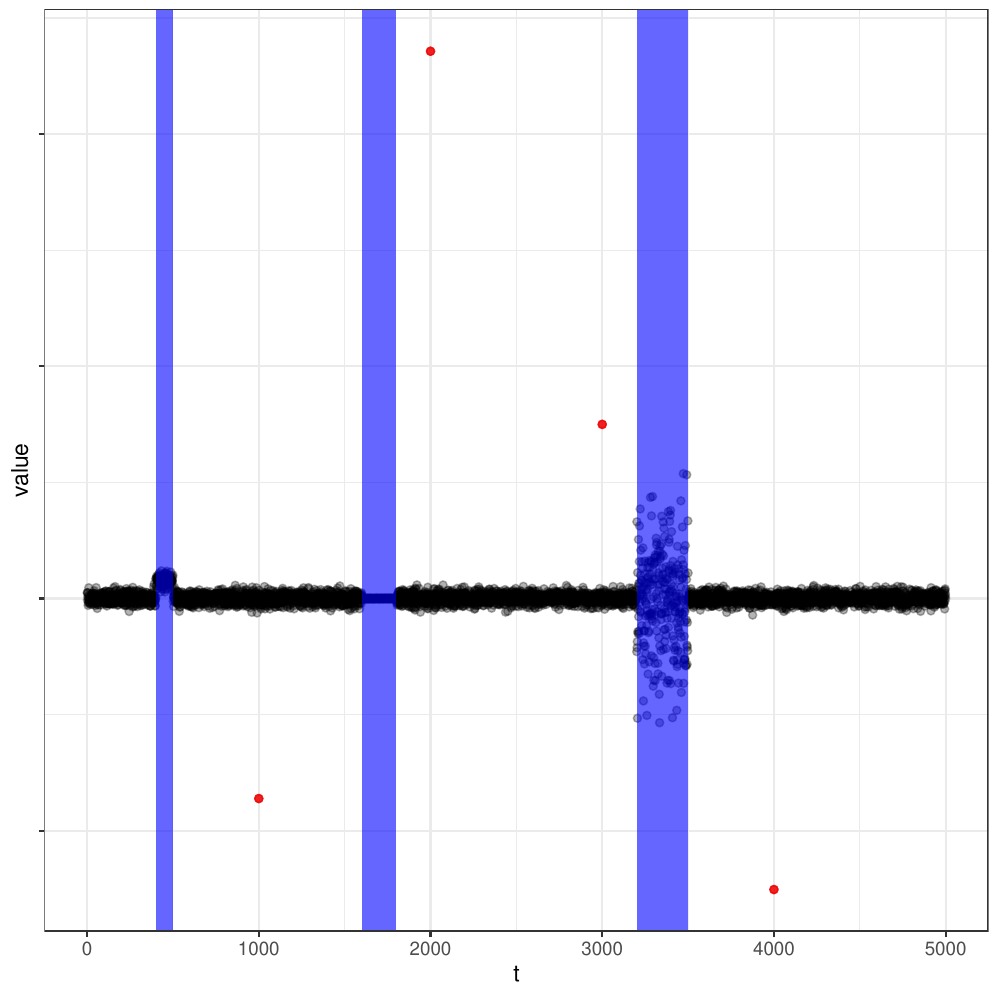}}
\end{minipage}%
\begin{minipage}{.5\linewidth}
\centering
\subfloat[]{\label{figure:uvcapa-simulated-data-2}\includegraphics[scale=.5]{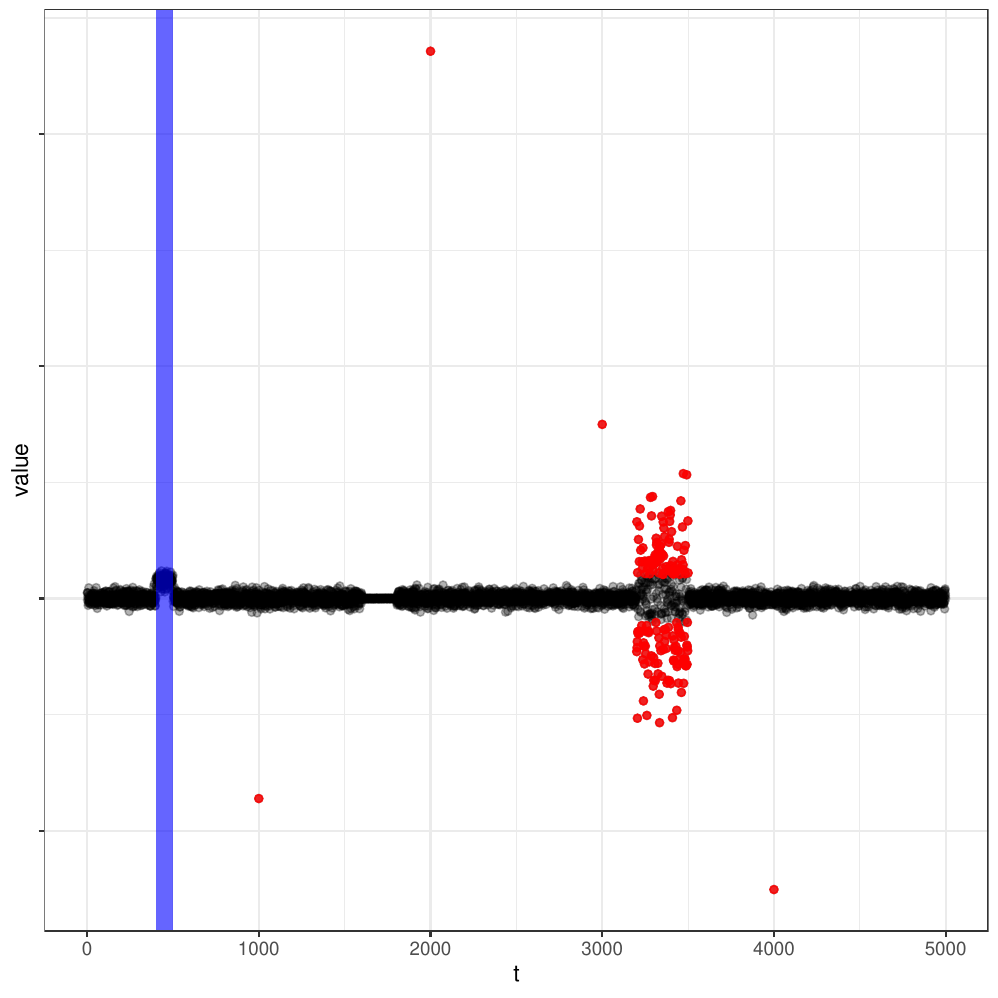}}
\end{minipage}%
\caption{Visualisation of data, collective and point anomalies detected by CAPA for the simulated univariate data.
(a) detecting changes in mean and variance. (b) detecting changes in mean only. The blue regions correspond to the predicted anomalous segments and the red dots indicate point anomalies.}
\label{fig:uvcapa-simulated-examples}
\end{figure}

To demonstrate univariate \code{capa} a data series of 5000 normally distributed observations with 3 collective anomalies and four point anomalies is analysed. The data
can be reproduced using the code provided below, which also runs the analyses and summarises the results.
\begin{CodeChunk}
\begin{CodeInput}
R> library("anomaly")
R> set.seed(0)
R> x <- rnorm(5000)
R> x[401:500] <- rnorm(100, 4, 1)
R> x[1601:1800] <- rnorm(200, 0, 0.01)
R> x[3201:3500] <- rnorm(300, 0, 10)
R> x[c(1000, 2000, 3000, 4000)] <- rnorm(4, 0, 100)
R> x <- (x - median(x)) / mad(x)
R> res <- capa(x)
R> summary(res)
\end{CodeInput}
\begin{CodeOutput}
Univariate CAPA detecting changes in mean and variance.
observations = 5000
minimum segment length = 10
maximum segment length = 5000

Point anomalies detected : 4
  location variate  strength
1     1000       1  43.07885
2     2000       1 117.84647
3     3000       1  37.49265
4     4000       1  62.67104

Collective anomalies detected : 3
  start  end variate start.lag end.lag  mean.change variance.change
1   401  500       1         0       0 14.597971638    4.990295e-04
2  1601 1800       1         0       0  0.001502774    9.869876e+01
3  3201 3500       1         0       0  0.036926415    7.764414e+00
\end{CodeOutput}
\begin{CodeInput}
R> plot(res)  
\end{CodeInput}
\end{CodeChunk}
The \code{summary} method displays information regarding the
analysis and details regarding the location and nature of the detected anomalies. The formatting demonstrates that \code{capa} correctly determines the presence of the
anomalies in the simulated data. The \code{plot} function generates a {\bf ggplot} object \citep{h-wickham-ggplot2-book} which is shown in Figure \ref{figure:uvcapa-simulated-data-1}. The location of the collective
anomalies are highlighted by vertical blue bands and the data point anomalies are shown in red.
By default, CAPA detects both changes in mean and variance. The option \code{type=''mean''} can be used to detect changes in mean only.
\begin{CodeChunk}
\begin{CodeInput}
R> res <- capa(x, type = "mean")
R> collective_anomalies(res)
\end{CodeInput}
\begin{CodeOutput}
    start end mean.change test.statistic
1   401 500    14.92774       1492.774    
\end{CodeOutput}
\begin{CodeInput}
R> head(point_anomalies(res))
\end{CodeInput}
\begin{CodeOutput}
  location  strength
1     1000  43.07885
2     2000 117.84647
3     3000  37.49265
4     3201  11.44038
5     3202  16.52037
6     3203  10.58874
\end{CodeOutput}
\end{CodeChunk}
In this case, \code{capa} correctly identifies the collective change in mean and the point anomalies. However, as a consequence of CAPA now looking for changes in mean only, and assuming constant variance, the analysis results in changes in variance being classified as groups of point anomalies, see Figure (\ref{figure:uvcapa-simulated-data-2}). The above example also demonstrates the \code{collective_anomalies} function,
which is used to produce a data frame containing the location and change in mean for collective anomalies, and the \code{point_anomalies} function which provides the location 
and strength of the point anomalies.

As previously noted, the CAPA algorithm assumes that the data has been standarised. When this is not the case, false anomalous regions may be identified, as is the case in the following example.
\begin{CodeChunk}
\begin{CodeInput}
R> res <- capa(1 + 2 * x, type = "mean")
R> nrow(collective_anomalies(res))
\end{CodeInput}
\begin{CodeOutput}
47
\end{CodeOutput}
\end{CodeChunk}
\subsubsection{Real data - machine temperature}  
\label{subsubsection:uvcapa-machine-temperature-data}
To demonstrate the application of \code{capa} to real univariate data, a data stream from the Numenta Anomaly Benchmark corpus \citep{AHMAD2017134} 
consisting of temperature sensor data of an internal component of a large industrial machine is analysed. The dataset is included, with permission, in the \textbf{anomaly} package on the condition that derived work be kindly requested to acknowledge \citep{AHMAD2017134}.

\begin{figure}
\begin{minipage}{.5\linewidth}
\centering
\subfloat[]{\label{fig:machine-temperature-example-a}\includegraphics[scale=.5]{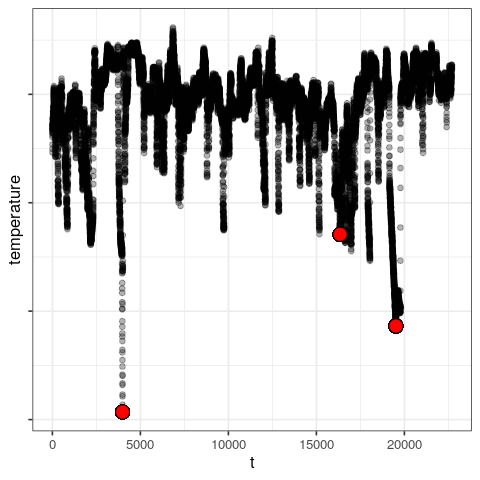}}
\end{minipage}%
\begin{minipage}{.5\linewidth}
\centering
\subfloat[]{\label{fig:machine-temperature-example-b}\includegraphics[scale=.5]{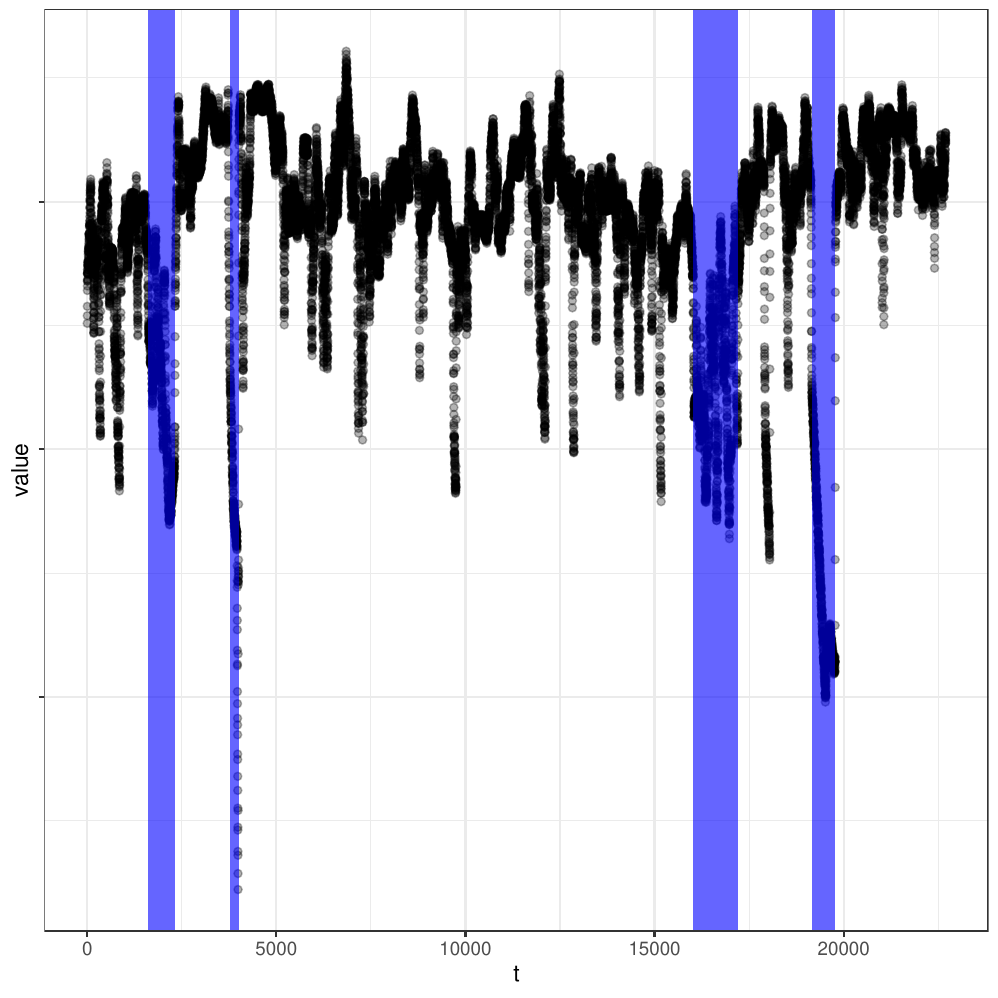}}
\end{minipage}\par\medskip

\caption{Time series of machine temperature data. 
(a) 
The highlighted data points show the locations of the anomalies identified by an engineer working on the machine.
(b) The anomalies identified by {\bf capa}. The second, third and fourth anomalies correspond well with the known anomalies.
The first anomaly may be a false positive or an anomaly not recorded by the engineer.}
\label{fig:machine-temperature-example}
\end{figure}

The machine temperature data consists of 22695 observations recorded at 5 minute intervals and contains three known anomalies as identified by an engineer working on the
machine (Figure \ref{fig:machine-temperature-example-a}). 
The first anomaly corresponds to a planned shutdown of the machine and the third anomaly to a catastrophic failure of the machine. The second anomaly, which can be difficult to
detect, corresponds to the onset of a problem which led to the eventual system failure \citep{DBLP:journals/corr/LavinA15}.  
Using \code{capa} with default parameters for the (normalised) data results in the detection of $97$ collective anomalies.  
\begin{CodeChunk}
\begin{CodeInput}
data("machinetemp")
attach(machinetemp)
x <- (temperature - median(temperature)) / mad(temperature)
res <- capa(x, type = "mean")
canoms <- collective_anomalies(res)
dim(canoms)[1]
\end{CodeInput}
\begin{CodeOutput}
[1] 97
\end{CodeOutput}
\end{CodeChunk}
One potential source of this over sensitivity is the presence of autocorrelation in the data. A robust estimate
for the $AR(1)$-autocorrelation $\hat{\rho}$ can be obtained using the \code{covMcd} method from the \pkg{robustbase} package.
\begin{CodeChunk}
\begin{CodeInput}
R> library("robustbase")
R> n <- length(x)
R> x.lagged <- matrix(c(x[1:(n - 1)], x[2:n]), n - 1, 2)
R> rho_hat <- covMcd(x.lagged, cor = TRUE)$cor[1,2]
\end{CodeInput}
\end{CodeChunk}
which gives an estimate for $\hat{\rho}$ of 0.987.

As mentioned above, the default penalties have a tendency to return many false positives on structured, i.e., non independent, data. Instead we can inflate the penalties based on our estimates of the autocorrelation.
\begin{CodeChunk}
\begin{CodeInput}
R> inflated_penalty <- 3 * (1 + rho_hat) / (1 - rho_hat) * log(n)
R> res <- capa(x, type = "mean", beta = inflated_penalty,
+    beta_tilde = inflated_penalty)
R> summary(res)
\end{CodeInput}
\begin{CodeOutput}
Univariate CAPA detecting changes in mean.
observations = 22695
minimum segment length = 10
maximum segment length = 22695

Point anomalies detected : 0

Collective anomalies detected : 4
  start   end variate start.lag end.lag mean.change test.statistic
1  1612  2327       1         0       0    9.148952       6550.650
2  3773  4002       1         0       0   25.648888       5899.244
3 16023 17204       1         0       0    8.191733       9682.628
4 19166 19775       1         0       0   39.426847      24050.377
\end{CodeOutput}
\begin{CodeInput}
R> plot(res)
\end{CodeInput}
\end{CodeChunk}
The new predicted collective anomalies are shown in Figure \ref{fig:machine-temperature-example-b}. The second, third and fourth  collective anomalies detected using the modified
penalty values correspond well with the known anomalies. The first anomaly on the other hand does not have a corresponding label which means that it is either a false positive or an anomaly corresponding to an event which has not been detected or recorded. 
Note that the test statistic for the first anomaly is stronger than for the second, but has smaller change in mean. This is inherent in the definition for the value of the test
statistic used when inferring changes in mean, which is the change in mean multiplied by the length (duration) of the anomaly.

\subsection[Multivariate CAPA]{Multivariate CAPA} \label{subsection:mcapa}
The \code{capa} function also has provision for analysing multivariate data series using a multivariate version of the CAPA algorithm \citep{2019arXiv190901691F}. The algorithm assumes that the $p$ components of the time series are independent of one another in all aspects except the locations of collective anomalies, which can affect any subset of the components. As with the univariate case, the current implementation of CAPA assumes non-anomalous data is independent normally distributed with a component specific mean and variance. Anomalous regions are then regions with a different mean, or a different mean and variance.
The saving for a collective anomaly starting at $s$ and ending at $e$ involves aggregating the savings across components
\begin{equation*}
S(s,e) = \max_k \sum_{i=1}^k \left( S_{(i)}(s,e) - \beta_i  \right)
\end{equation*}
Here, $S_{(1)}(s,e) \geq ... \geq S_{(p)}(s,e) $ corresponds to the order statistics of the savings 
$S_{1}(s,e) , ... , S_{p}(s,e) $, with $S_{i}(s,e)$ denoting the improvement in the individual components, as defined in section \ref{section:Background}. The $\beta_i$ denote the typically decreasing marginal penalties or thresholds controlling false positives. Crucially, CAPA allows for the alignment of collective anomalies across components to be imperfect. In other words, certain components can lag by entering the anomalous state later and/or returning to their typical state earlier than others.

The (multivariate specific) \code{max_lag} argument in the \code{capa} function is used to set a limit on how much a collective anomaly in one variate can lag (or lead) a collective anomaly in another variate, whilst still being 
part of the same multivariate anomaly. The run time scales linearly with \code{max_lag}, though this dependence tends to be weak for small values of \code{max_lag}. The run time also scales linearly (up to logarithmic factors) with the number of components $p$. 
The default penalties are specific to i.i.d.\ data and tend to return many false positives when some of the $p$ series contain, for example, auto-correlated structure.  Extending the argument of \citet{LavielleMarc2000LEoa} to the multivariate setting, using
\begin{equation*}
\tilde{\beta} = 2 \frac{1+\hat{\rho}_1}{1-\hat{\rho}_1}\log(np) \;\;\;\;\;\;\; \beta_1 = 2 \frac{1+\hat{\rho}_1}{1-\hat{\rho}_1}\log(np(w+1))    \;\;\;\;\;\;\; \beta_i = 2 \frac{1+\hat{\rho}_i}{1-\hat{\rho}_i}\log(p(w+1))  \;\;\;\;\;\; 2 \leq i \leq p
\end{equation*}
can achieve good false positive control. Here, $\hat{\rho}_i$ is the $i$th largest of the robust estimates for the $AR(1)$-auto-correlation coefficients of the $p$ series and $w$ the maxlag.
\subsubsection{Simulated data 1}
\label{subsubsection:example-3-mvcapa-simulated-data-1}
\begin{figure}
\begin{minipage}{0.5\linewidth}
\centering
\subfloat[]{\label{fig:mvcapa-simulated-example}\includegraphics[scale=0.5]{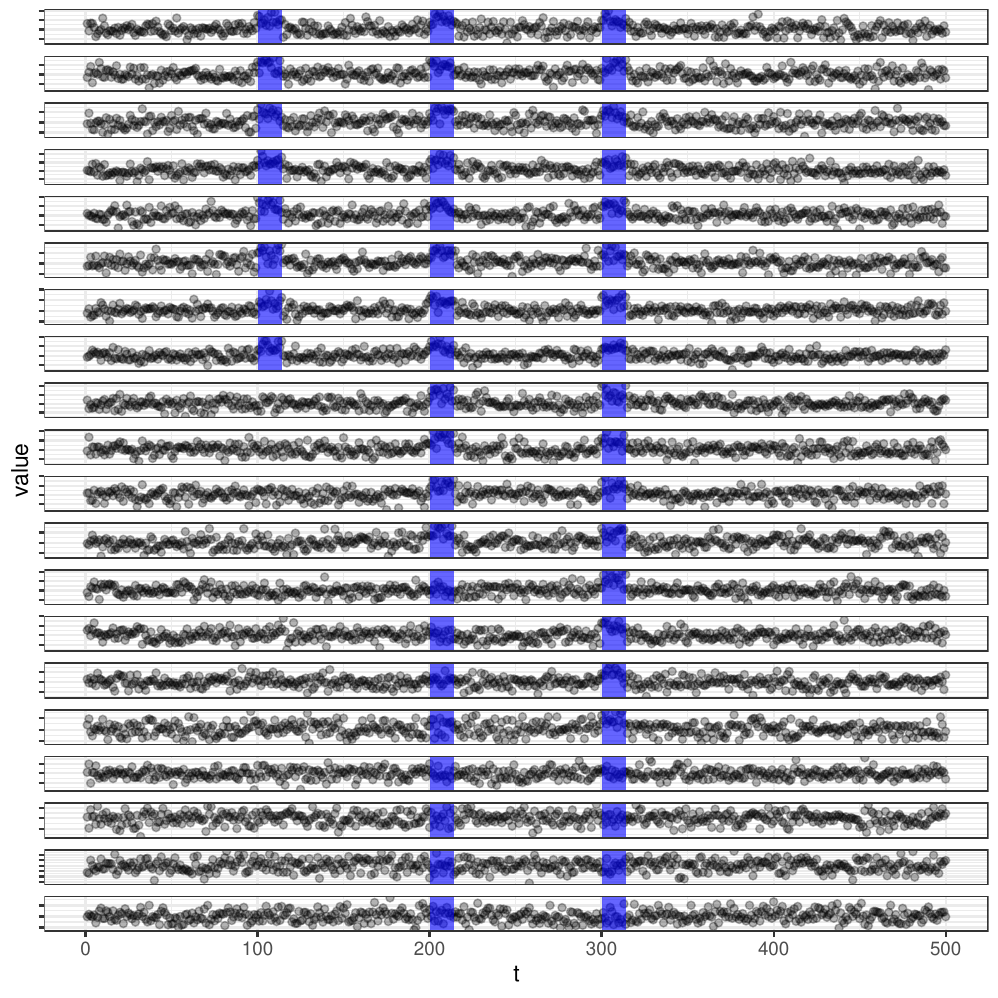}}
\end{minipage}%
\begin{minipage}{.5\linewidth}
\centering
\subfloat[]{\label{fig:mvcapa-simulated-example-fp-control}\includegraphics[scale=.5]{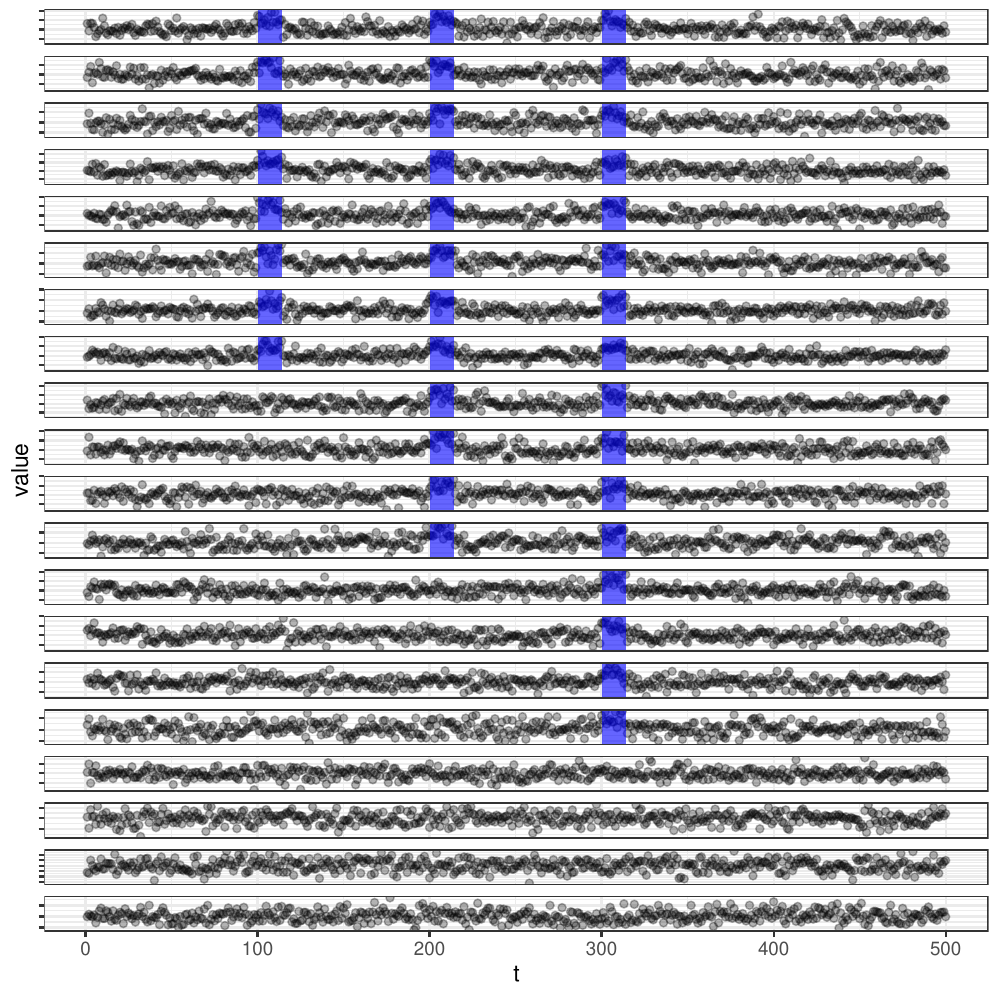}}
\end{minipage}%
\caption{CAPA results for Simulated data 1 in Section \ref{subsection:mcapa}. The variates are displayed in order from the top down and the location and extent of the anomalous regions are indicated by the blue areas.
  (a) Using a penalty which maximises power for the detection of collective anomalies but does not control false positives in the components.
  (b) Using a penalty which controls false positives in the components (at a loss of power).
}
\label{fig:mvcapa-simulated-example-components}
\end{figure}

To demonstrate multivariate CAPA, a simulated data set, \code{sim.data}, consisting of 500 observations on 200 variates which are $\mathcal{N}(0,1)$ distributed is used. The data is provided by the \pkg{anomaly} package
and contains three multivariate anomalies of length 15 located at $t=100$, $t=200$, and $t=300$ for which the mean changes from 0 to 2. The anomalies affect variates 1 to 8, 1 to 12 and 1 to 16
respectively. 
Figure \ref{fig:mvcapa-simulated-example} shows a tile plot of the data and the anomaly locations as estimated by the following analysis. 

\begin{CodeChunk}
  \begin{CodeInput}
R> data("simulated")
R> res <- capa(sim.data, type = "mean", min_seg_len = 2)
R> plot(res, subset = 1:20)
\end{CodeInput}
\end{CodeChunk}
%


Clearly the overall positions of the anomalies have been located correctly however, many false positive anomalous segments have been fitted across most of the variates. This issue arises because the default penalty used by \code{capa} is tuned towards detection accuracy at the expense of false positive control in the number of components fitted as anomalous. False positive control can be recovered, at a loss of power against anomalies weakly affecting a lot of components, by using regime 2 from \cite{2019arXiv190901691F}:
%
\begin{CodeChunk}
\begin{CodeInput}
R> beta <- 2 * log(ncol(sim.data):1)
R> beta[1] <- beta[1] + 3 * log(nrow(sim.data))
R> res <- capa(sim.data, type= "mean", min_seg_len = 2,beta = beta)
R> plot(res, subset = 1:20)
\end{CodeInput}
\end{CodeChunk}
As is apparent from Figure \ref{fig:mvcapa-simulated-example-fp-control}, CAPA now controls false positives. Unfortunately, in general, optimal power and false positive control in the number of variates cannot both be achieved, as shown by \cite{tony2011optimal}. 
%
%
%
%
%
\subsubsection{Simulated data 2}
\label{subsubsection:example-3-mvcapa-simulated-data-2}
\begin{figure}
\centering
\includegraphics{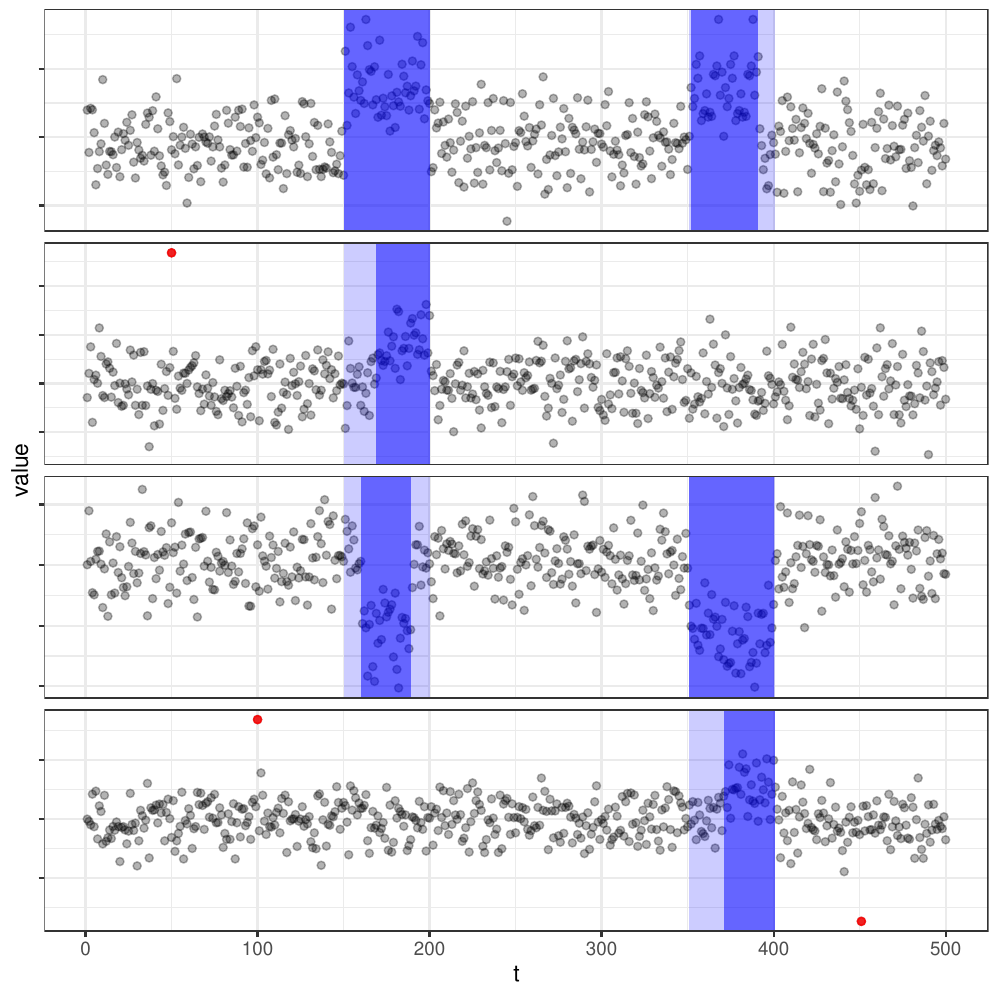}
\caption{CAPA output on simulated data in Section \ref{subsubsection:example-3-mvcapa-simulated-data-2}. Collective anomalies are coloured in blue, with lags displayed in light blue.}
\label{fig:MVCAPAex2}
\end{figure}

As mentioned previously, a maximum lag can be used when it is suspected that the collective anomalies do not perfectly align. This requires minor modifications to the argument structure 
\begin{CodeChunk}
\begin{CodeInput}
R> set.seed(0)
R> x1 <- rnorm(500)
R> x2 <- rnorm(500)
R> x3 <- rnorm(500)
R> x4 <- rnorm(500)
R> x1[151:200] <- x1[151:200] + 2
R> x2[171:200] <- x2[171:200] + 2
R> x3[161:190] <- x3[161:190] - 3
R> x1[351:390] <- x1[371:390] + 2
R> x3[351:400] <- x3[351:400] - 3
R> x4[371:400] <- x4[371:400] + 2
R> x4[451] <- x4[451] * max(1, abs(1 / x4[451])) * 6
R> x4[100] <- x4[100] * max(1, abs(1 / x4[100])) * 6
R> x2[050] <- x2[050] * max(1, abs(1 / x2[050])) * 6
R> x1 <- (x1 - median(x1)) / mad(x1)
R> x2 <- (x2 - median(x2)) / mad(x2)
R> x3 <- (x3 - median(x3)) / mad(x3)
R> x4 <- (x4 - median(x4)) / mad(x4)
R> x <- cbind(x1, x2, x3, x4)
R> res <- capa(x, max_lag = 20, type = "mean")
R> plot(res)
\end{CodeInput}
\end{CodeChunk}
The output of this analysis can be found in Figure \ref{fig:MVCAPAex2}.

\section[Proportion Adaptive Segment Selection (PASS)]{Proportion Adaptive Segment Selection (PASS)}
\label{section:pass}
The \pkg{anomaly} package includes a scalable implementation of the Proportion Adaptive Segment Selection (PASS) algorithm by \cite{10.1093-biomet-ass059}.   
PASS is an algorithm designed to efficiently scan long multi-variate sequences of data using a test statistic that detects multivariate collective anomalies in mean and/or variance. For each candidate collective anomaly, with a given start point and end point, $s$ and $e$ say, PASS tests each component individually for a mean anomaly thus obtaining $p$-values $q_1,...,q_p$. These component specific $p$-values are then ordered $q_{(1)} \leq ... \leq q_{(p)}$ and combined into a test-statistic, $q$, for the segment $(s,e)$ that combines information across the $p$-values using higher criticism \citep{donoho2004}:  
\begin{equation} \label{eq:PASSq}
q = \max_{\alpha_0 \leq i \leq p}  \left( \sqrt{p} \frac{\frac{i}{p} - q_{(i)}}{\sqrt{q_{(i)}(1 - q_{(i)}) }} \right) 
\end{equation}
for an integer $\alpha_{0} \geq 1$. To fit a multiple collective anomalies, an algorithm similar to Circular Binary 
Segmentation \citep{10.1093/biostatistics/kxh008},  is used.

PASS inherits most of its hyper parameters and properties from higher criticism. In particular, it is often suggested to set $\alpha_{0} > 1$, i.e., to disregard some of the lowest $p$-values when using higher criticism  to stabilise the 
procedure. However, this can lead to anomalies affecting fewer than $\alpha_{0}$ components escaping detection. Furthermore, their approach requires selecting a suitable threshold 
value $\lambda$, which is typically increased with the data dimension $n$ and $p$. Low values of $\alpha_{0}$ can also make inflation of $\lambda$ advisable; guidance on which is given 
in \citet[Section~3.1]{10.1093-biomet-ass059}. The method has been implemented in compiled code by following
steps 1 to 8 in \citet[section~2.2]{10.1093-biomet-ass059} and has computational complexity $\mathcal{O}(max\_seg\_len \cdot np\log(p))$. 

The \pkg{anomaly} package provides the function \code{pass} which accepts the following arguments :
\begin{itemize}
\item \code{x} - A numeric matrix with n rows and p columns containing the data which is to be inspected. The time series data classes ts, xts, and zoo are also supported.
\item \code{alpha} - An integer value greater then 0 corresponding to $\alpha_{0}$ in \citet{10.1093-biomet-ass059}.
This value is used to stabilise the higher criticism based test statistic used by PASS leading to a better finite sample familywise error rate.  Anomalies
affecting fewer than alpha components will be more likely to escape detection. The default value is 2.
\item \code{lambda} - A positive real value setting the threshold value for the familywise Type 1 error.
The default value is $(1.1 {\rm log}(n \times \rm{max\_seg\_len)} +2 {\rm log}({\rm log}(p))) / \sqrt{{\rm log}({\rm log}(p))}$. 
\item \code{max_seg_len} -  A positive integer corresponding to the maximum segment length. This parameter corresponds to the maximum interval length, $L$, in \cite{10.1093-biomet-ass059}. 
The default value is 10.
\item \code{min_seg_len} -  A positive integer (\code{max_seg_len} >= \code{min_seg_len} > 0) corresponding to the minimum segment length. The default value is 1. 
\end{itemize}
\subsection{PASS - Simulated example}
The following code demonstrates how the \code{pass} method provided by \pkg{anomaly} is used. In this example the data, \code{sim.data}, is the same as that used in the
\code{capa} example in Section \ref{subsubsection:example-3-mvcapa-simulated-data-1}
\begin{CodeChunk}
\begin{CodeInput}
R> library("anomaly")
R> data("simulated")
R> res <- pass(sim.data, max_seg_len = 20, alpha = 3)
R> collective_anomalies(res)
\end{CodeInput}
\begin{CodeOutput}
  start end      xstar
1   200 214 1519317784
2   100 114   42907782
3   300 315   22296743
\end{CodeOutput}
\end{CodeChunk}
The results show the start and end of each anomaly along with \texttt{xstar} denoting the value of the higher criticism test statistic for the segment, that is the value of $q$ as defined by (\ref{eq:PASSq}). Larger values indicate more evidence for a collective anomaly, and the segments are listed in decreasing order of \texttt{xstar}. The results are consistent with those provided by \code{capa} in that the three anomalies are all detected.
However, unlike MVCAPA, PASS does not indicate which series are anomalous.

%
%
\section[Bayesian Abnormal Region Detector (BARD)]{Bayesian Abnormal Region Detector (BARD)} \label{section:bard}

The Bayesian Abnormal Region Detector (BARD) \citep{bardwell2017} is a fully Bayesian method for estimating abnormal regions in multivariate data. It assumes that data has been normalised so that data for each variate in a normal region has mean 0 and variance 1, and that abnormal regions correspond to a change in mean. Specifically, the model is a special case of (2), where the parameter, $\mathbf{\theta}^{(i)}(t)$ is the mean of variate $i$ at time point $t$, and we model that, conditional on the parameters, the data are independent Gaussian. 

As it is a Bayesian approach, BARD differs from CAPA and PASS in two aspects. First, the user has to specify prior distributions for aspects of the model such as the mean in abnormal segments, and the length of normal and abnormal segments. Second, the output of the algorithm will be draws from a posterior distribution, which can be used to produce a single estimate of the location of the abnormal segments or give some measure of uncertainty about where the abnormal segments are. Like PASS, BARD only gives information about where the abnormal segments are located and not which variates are abnormal within each segment. 

The parametric form of the prior distributions assumed by BARD are as follows. Segment lengths are assumed to have a negative binomial distribution, with parameters $(k_N,p_N)$ for normal segments and $(k_A,p_A)$ for abnormal segments, where a negative binomial random variable with parameters $(k,p)$ has probability mass function
\[
\Pr(X=x)=  \left( \begin{array}{c} x + k - 1 \\  x \end{array} \right) (1-p)^k p^x,
\]
with $\mbox{E}(X)= \frac{k(1-p)}{p}$ and $\mbox{Var}(X)= \frac{k(1-p)}{p^2}$.

For an abnormal segment we need to further define a prior for the segment mean, $\mu$, and this is assumed to uniform on a range for $|\mu|$, with the sign of the mean being equally likely to be positive or negative. We also need to specify the average proportion of variates affected by an abnormal segment, and the probability that an abnormal segment is followed by a further abnormal segment.

The BARD algorithm proceeds in two stages. First it calculates an approximation to the joint posterior distribution for the number and location of the abnormal segments. The approximation comes first from using numerical integration to calculate marginal likelihoods, and second from using probabilistic pruning (also known as resampling) within a particle filter to ensure the algorithm's complexity is linear in the number of time points. These parts of the algorithm can be controlled by the user, but empirical evidence in \cite{bardwell2017} suggest that the approximation error when using the default choices is small.

The second step of BARD is to draw a number of independent samples from the posterior. The individual draws can either be plotted to give a sense of the uncertainty around where the abnormal segments are, or can be summarised by a single point estimate of their location. The anomaly package provides functions to do both of these. The approach taken to summarise the posterior by a single point estimate is to consider marginally each time-point, $t$, and the proportion of draws which place $t$ within an abnormal segment. Our point-estimate flags point $t$ as within an abnormal segment if and only if this proportion of draws is above some user-chosen threshold. See Figure \ref{fig:bard-marginal-prob-plot} for an example.

The \pkg{anomaly} package provides the function \code{bard} which accepts the following arguments :
\begin{itemize}
\item \code{x} - A numeric matrix with n rows and p columns containing the data which is to be inspected. The time series data classes ts, xts, and zoo are also supported.
\item \code{p\_N} - Hyper-parameter of the negative binomial distribution for the length of non-anomalous segments (probability of success). Defaults to $\frac{1}{n+1}$.
\item \code{p\_A} - Hyper-parameter of the negative binomial distribution for the length of anomalous segments (probability of success). Defaults to $\frac{5}{n}$.
\item \code{k\_N} - Hyper-parameter of the negative binomial distribution for the length of non-anomalous segments (size). Defaults to 1.
\item \code{k\_A} - Hyper-parameter of the negative binomial distribution for the length of anomalous segments (size). Defaults to $\frac{5p_A}{1- p_A}$.
\item \code{pi\_N} - Probability that an anomalous segment is followed by a non-anomalous segment. Defaults to 0.9.
\item \code{paffected} -  Proportion of the variates believed to be affected by any given anomalous segment. Defaults to 5\%. 
This parameter is relatively robust to being mis-specified and is studied empirically in Section 5.1 of \cite{bardwell2017}.
\item \code{lower} - The lower limit of the the prior uniform distribution for the mean of an anomalous segment $\mu$. Defaults to $2\sqrt{\frac{\log(n)}{n}}$.
\item \code{upper} - The upper limit of the prior uniform distribution for the mean of an anomalous segment $\mu$. 
Defaults to the largest value of x.
\item \code{alpha} -  Threshold used to control the resampling in the approximation of the posterior distribution at each time step. A sensible default is 1e-4. Decreasing alpha increases the accuracy of the posterior distribution but also increases the computational complexity of the algorithm.
\item \code{h} - The step size in the numerical integration used to find the marginal likelihood. The quadrature points are located from \code{lower} to \code{upper} in steps of \code{h}. Defaults to 0.25. Decreasing this parameter increases the accuracy of the calculation for the marginal likelihood but increases computational complexity.
\end{itemize}
\subsection{BARD - Simulated example}
The following code demonstrates how the \code{bard} method provided by \pkg{anomaly} can be used. In this example the data, \code{sim.data}, is the same as that used in Section 4, with PASS.
%
%
\begin{CodeChunk}
\begin{CodeInput}
R> library("anomaly")
R> data("simulated")
R> bard.res <- bard(sim.data)
\end{CodeInput}
\end{CodeChunk}
The priors (\code{p\_N},\code{k\_N},\code{p\_A} and \code{k\_A}) for the two length of stay distributions for normal and abnormal segments were chosen to be quite vague but 
with abnormal segments being much smaller than their normal counterparts. The mean (standard deviation) for normal segments is 190 (62) whereas for abnormal segments it is 10 (4).
With no particular knowledge of the process in question we took the probability that an abnormal segment (\code{pi\_N}) is followed by a normal segment as 90\%. 
This was relatively arbitrary and assigned a high prior probability to the classic epidemic changepoint model but still allows for two abnormal segments to follow each 
other (albeit in different variates). The proportion of variates assumed to be affected by an abnormal segment (\code{paffected})was taken to be 5\% of the total number of variates. This 
proportion is small enough to be able to locate rare anomalies.  The prior for the mean $\mu \sim \mathcal{U}(\mathbf{lower}, \mathbf{upper})$ was taken to be in the range of 0.5 to 1.5. The lower 
limit of 0.5 gives the minimum change in mean we are interested in detecting. To calculate the marginal likelihood of abnormal segments numerical integration was used with a step 
size (\code{h}) set at 0.25. In the example, the default value for the threshold parameter $\alpha$ of $1 \times 10^{-4}$ has been used.   

The \code{bard} function returns an S4 class that includes the posterior distribution of the abnormal segments given the observed data. To obtain samples from the posterior, and, from these, posterior estimates
for the location collective anomalies the \code{sampler} function is used.
\begin{CodeChunk}
\begin{CodeInput}
R> sampler.res <- sampler(bard.res, gamma = 1/3, num_draws = 1000)
R> show(sampler.res)
\end{CodeInput}
\begin{CodeOutput}
BARD sampler detecting changes in mean
observations = 500
variates = 200
Collective anomalies detected : 3
  start end LogMargLike
3   199 213    319.8889
4   299 313    311.7028
2    99 113    177.8095
\end{CodeOutput}
\begin{CodeInput}
R> plot(sampler.res, marginals = TRUE)
\end{CodeInput}
\end{CodeChunk}
A number of samples (\code{num_draws}) are taken from the posterior, and from these estimates of the location of the collective anomalies are obtained based on the asymmetric loss  \citep{bardwell2017} using the parameter \code{gamma}. This loss will estimate a location $t$ as part of a collective anomaly if the proportion of posterior samples that have $t$ in a collective anomaly is greater than $1/(1+\gamma)$.
The show function reports the resulting estimated collective anomalies, together with a measure of the evidence (LogMargLike) for the collective anomaly in terms of the log marginal likelihood for the region being a collective anomaly rather than a part of a normal region. Larger values imply a stronger anomaly, which is dependent on the length, change magnitude and number of affected variates of each collective anomaly. The example shows that the location and  
relative strength of the three abnormal segments are broadly consistent with those obtained using \code{capa} and \code{pass}.

\begin{figure}
\centering
\includegraphics[scale=0.4]{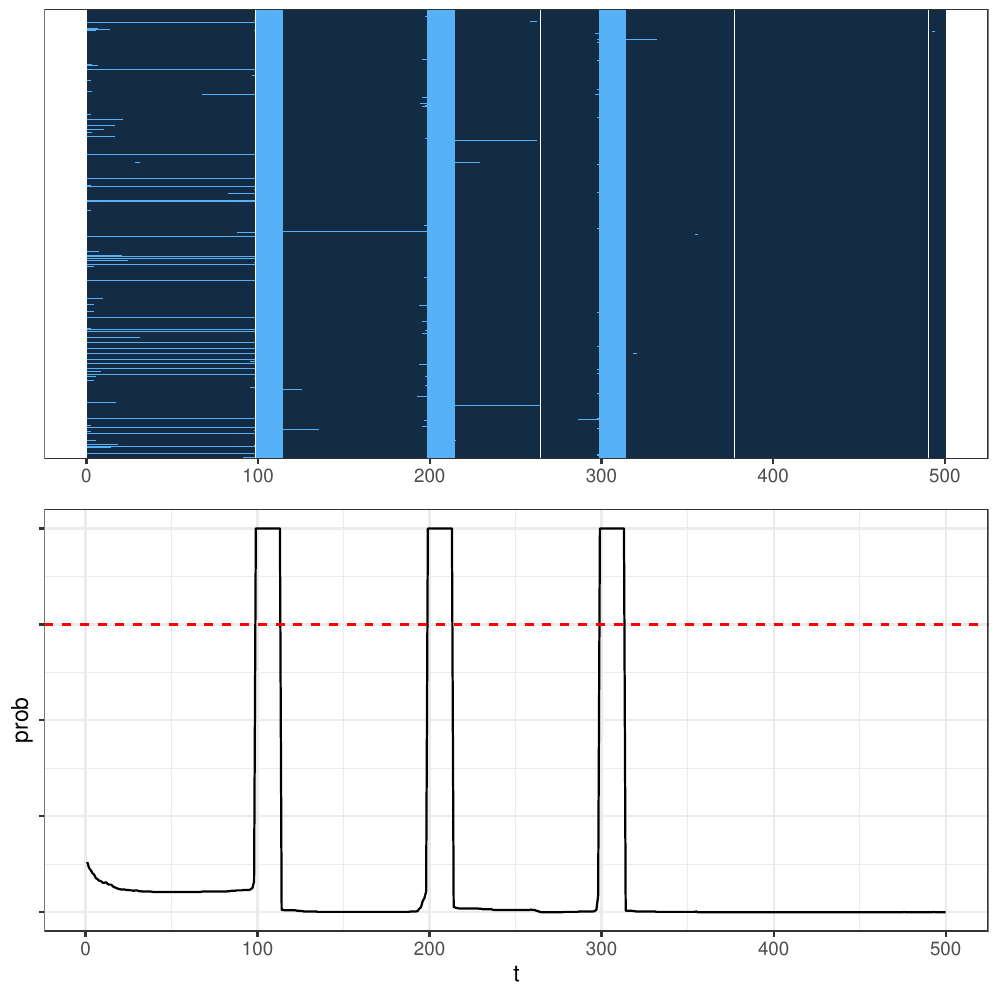}
\caption{Plot of a sample of 1000 draws from the posterior distribution (top), each realisation is shown in a separate row with light blue regions indicating the time points in that realisation have been inferred as part of a collective anomaly. The marginal probability of a collective anomaly (bottom), with the user defined threshold (for $\gamma=1/3$) for flagging a region as a collective anomaly (horizontal dashed line).}
\label{fig:bard-marginal-prob-plot}
\end{figure}

As with \code{capa} and \code{pass}, the \code{plot} function can be used to visualise the data. However, the argument \code{marginals} can be used to display additional information
as shown in Figure \ref{fig:bard-marginal-prob-plot}. The top plot shows different realisations from the posterior distribution. The values are either 0 or 1 indicating if the time point in the realisation is part of
a collective anomaly or not. 
The marginal probability of each time point being a collective anomaly is shown in the bottom plot. This is the fraction of sampled realisations that were found to be anomalous at each time.
The dashed horizontal line is the threshold, for our choice of \code{gamma=1/3}, for which a collective anomaly is inferred. 

\section{CAPA and PASS - Micro array data}  
\label{subsection:mvcapa-pass-micro-array-data}
\begin{figure}[ht]
\centering
\includegraphics[scale=.5]{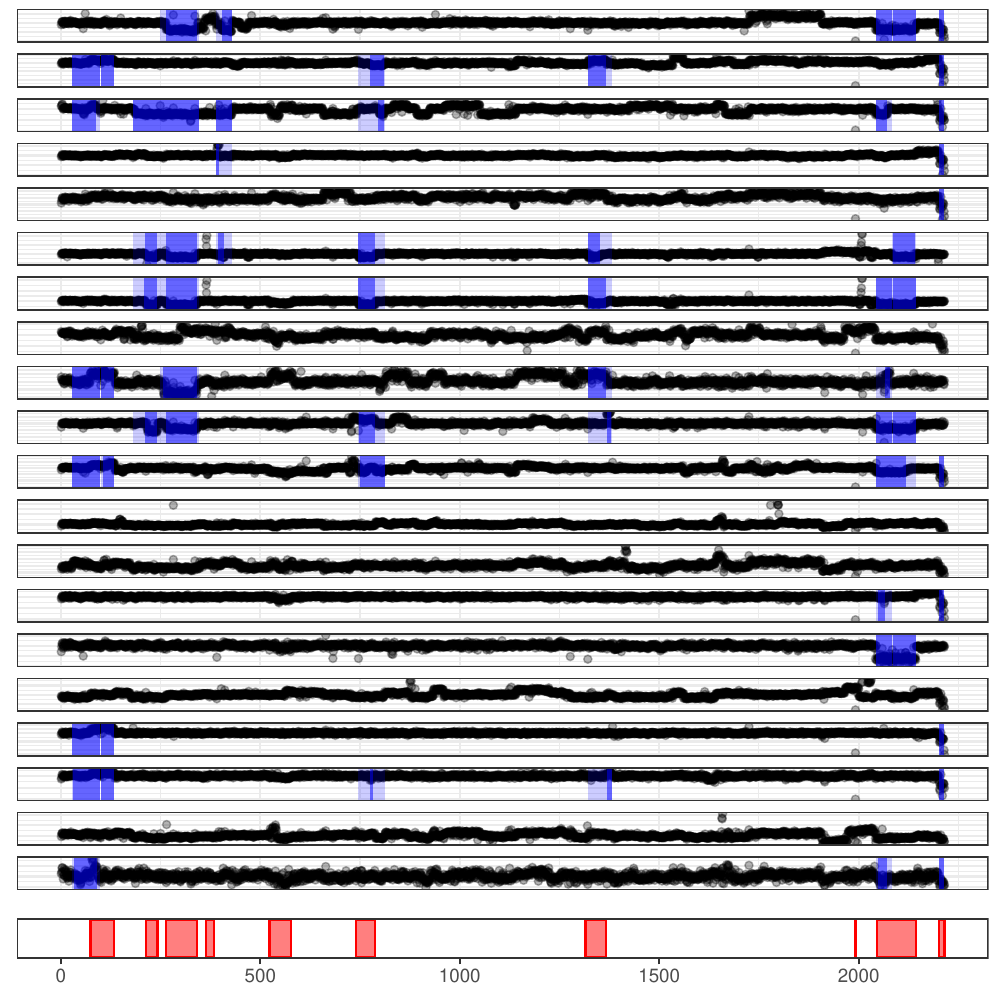}
\caption{Locations of the 10 strongest anomalies for \code{capa} (blue) and \code{pass} (red).}
\label{fig:mvcapa-acgh}
\end{figure}

%
%
This example examines microarray data for 20 individuals with a bladder tumour from the ACGH data set which is available in the \pkg{ecp} package \citep{ecp-package}. 
\begin{CodeChunk}
\begin{CodeInput}
R> library("ecp")
R> data("ACGH")
R> acgh <- ACGH[[1]][,1:20]
\end{CodeInput}
\end{CodeChunk}
The data is highly autocorrelated so we transform each individual variate using equation \ref{eqn:ac_corrected_transform} for both \code{capa} and \code{pass} to avoid false positives.
%
\begin{CodeChunk}
\begin{CodeInput}
R> ac_corrected <- function(X){
      n <- length(X)
      rcor <- covMcd(matrix(c(X[2:n], X[1:(n-1)]), ncol = 2), cor = TRUE)
      psi <- rcor$cor[1,2]
      correction_factor <- sqrt((1 - psi) / (1 + psi))
      return(correction_factor * (X - median(X)) / mad(X))
   }
R> acgh <- acgh |> data.frame() |> sapply(ac_corrected)
\end{CodeInput}
\end{CodeChunk}
The acgh data is analysed using \code{max_seg_len = 200} to ensure that detected anomalies can cover the whole length of a single chromosome and, for \code{capa}, the maximum lag was set to 50 to prevent adjacent anomalies overlapping. 
\begin{CodeChunk}
\begin{CodeInput}
R> res.capa <- capa(acgh, type = "mean", max_lag = 50, max_seg_len = 200)
R> res.pass <- pass(acgh, max_seg_len = 200, alpha = 3)
\end{CodeInput}
\end{CodeChunk}
Both methods find that most of the region is anomalous (at least relative to the mean-0 model) so, for clarity of explanation,  
only the locations of the strongest 10 collective anomalies for each method are considered. These are shown in Figure \ref{fig:mvcapa-acgh} which demonstrates that the two sets of predicted anomalies are well alligned. Note however, only CAPA can distinguish between individuals that are affected by a specific anomaly or not.
\section{Discussion}
The detection of anomalous points and regions within data sequences and time series is becoming an increasingly important in many fields, from astrophysics to digitial networking. The \pkg{anomaly} package implements a number of recently proposed, computationally efficient statistical approaches, accessible via a simple, easy to use \proglang{R} interface. \pkg{anomaly} provides a first implementation of the collective and point anomaly (CAPA) family of anomaly detection methods that can be used to detect both point anomalies in otherwise normal segments of data, as well as detecting collective anomalies. It also provides implementations of the BARD and PASS methods. Distinctive plot classes have also been developed, allowing for the clear differentiation of  anomaly types in both univariate and multivariate settings. 

Each of the introduced methods are founded on independence assumptions though as described, in practice, they may be adapted to handle some (moderate) auto-correlation. Extension of these approaches to more general, time-dependent settings is the subject of current research. We hope to make such methods available within the package in due course, together with recently developed methods that allow for cross-dependence between series \cite{Tveten2022}.  

Each of the introduced methods also assumes that non-anomalous data is drawn from an underlying data generating process where the mean and variance are constant. 
In practice, trend and seasonality may be present within time series of interest. In such settings one might consider detrending/deseasonalising the data prior to running, e.g., CAPA. We advise care with such pre-processing, to ensure that anomalies of interest are not distorted by the process.



\section*{Acknowledgments}
Fisch gratefully acknowledges funding provided by EPSRC and BT under the Industrial CASE scheme, and support from the EPSRC-funded STOR-i Centre for Doctoral Training. In addition, Eckley \& Fearnhead gratefully acknowledge the financial support of EPSRC and BT through grants EP/N031938/1 and EP/R004935/1.

Section \ref{subsubsection:uvcapa-machine-temperature-data} uses data which is publicly available from the Numenta Anomaly Benchmark corpus \citep{AHMAD2017134} which can be accessed at https://github.com/numenta/NAB. It is included, with permission, in the anomaly package. Please remember to acknowledge \cite{AHMAD2017134} when using the data.

\newpage
\bibliography{jss4257}

\end{document}